\documentstyle[twocolumn,aps,graphicx]{revtex}

\begin{document}

\hyphenation{Ka-pi-tul-nik}

\twocolumn[
\hsize\textwidth\columnwidth\hsize\csname@twocolumnfalse\endcsname
\draft

\title{Microcantilever Studies of Angular Field Dependence of Vortex Dynamics
in BSCCO}
\author{J. Chiaverini, K. Yasumura, and  A. Kapitulnik}
\address{Departments of Applied Physics and of Physics, Stanford
University, Stanford, CA 94305, USA}

\date{\today}
\maketitle

\begin{abstract}

Using a nanogram-sized single crystal of BSCCO attached to a
microcantilever we demonstrate in a direct way that in magnetic
fields nearly parallel to the {\it ab} plane the magnetic field
penetrates the sample in the form of Josephson vortices rather
than in the form of a tilted vortex lattice. We further
investigate the relation between the Josephson vortices and the
pancake vortices generated by the perpendicular field component.

\end{abstract}

 ]

Much progress has been made in the understanding of the phase
diagram of layered superconductors  with very weak interlayer
coupling such as Bi$_2$Sr$_2$CaCu$_2$O$_{8-x}$ (BSCCO). In a
magnetic field perpendicular to the Cu-O planes, vortex lines are
viewed as stacks of pancake vortices that are weakly coupled via
Josephson and magnetic interactions in adjacent layers. Depending
on the relative strength of the two interactions we expect
different scenarios for the melting and decoupling transitions.
With stronger Josephson coupling, pancake vortices may remain
coupled below the melting transition.  However, for the case with
stronger magnetic coupling, a ``sublimation'' transition is
predicted in which the three-dimensional vortex lattice
dissociates into independent two-dimensional pancake vortices at
the melting transition \cite{blatter1}.

The study of vortex structures and melting transitions in magnetic
fields canted with respect to the c-axis has received mostly
theoretical consideration \cite{bulaevskii1,benkraouda,feinberg}.
A rich variety of vortex configurations has been proposed to take
into account the finite angle between the applied field and the
crystal {\it c} axis, as well as the layered structure, each
derived in a different regime of the parameters. In particular, an
interesting structure has been proposed for the case of
predominantly magnetic coupling between the layers
\cite{koshelev1}. In this case the in-plane magnetic field
interacts with the pancake vortices only through the Josephson
interaction, while the alignment of the pancakes is determined by
the magnetic coupling. Since this implies that a rigid tilt of the
vortex lattice costs magnetic energy, it would be more favorable
for the magnetic field to penetrate into the superconductor in the
form of Josephson vortices between the layers. The consequence of
the above considerations is a lattice of pancake vortices that
coexists with a lattice of Josephson vortices. Depending on the
angle, one lattice will be more dilute than the other.

The melting of the vortex lattice in BSCCO single crystals in a
tilted magnetic field was studied recently by Schmidt {\it et
al.} \cite{schmidt} using local Hall probe measurements with an
active area of $80\times 80\mu m^2$. They  found agreement with
published results by defining an effective melting field, {\it
i.e.} the perpendicular component of the magnetic field. The most
striking evidence that the thermodynamics follows the
perpendicular component was the observation that the entropy jump
at the transition is insensitive to the presence of the in-plane
field. This result did not obtain, however, at large angles where
the field orientation is close to the {\it ab} plane. Contrary to
the Schmidt {\it et al.} results, Ooi {\it et al.} \cite{ooi},
also using local magnetization measurements, though with an active
sample area three times smaller, found that the simple
perpendicular-component scaling ceases to work at much higher
angles (as measured from the {\it c} axis), thus invoking a new
temperature dependent formula to scale the angle-dependent
location of the first order transition peak.

In this paper we present results on magneto-mechanical
measurements of a nanogram-sized BSCCO single crystal in a tilted
magnetic field as a function of temperature, employing a
microcantilever device. The size of the crystal was chosen to be
smaller than the active area of previously reported local
measurements \cite{schmidt,ooi,yamaguchi}, while in this
experiment the measurement itself is inherently global. Choosing
nominal magnetic fields above the first-order transition
\cite{zeldov}, we observe a flat dissipation and a resonant
frequency of the device very close to the normal state value.
Tilting the sample with respect to the magnetic field reduces the
perpendicular component of the field and the sample appears to
undergo a melting transition at a perpendicular component that is
very close to the values published in the literature for similar
crystals. At nearly parallel field we observe oscillations
corresponding to single Josephson vortices as they move into and
out of the sample. This demonstrates for the first time that even
a small in-plane field does not cant the vortex lattice but rather
penetrates the superconductor in the form of Josephson vortices.

A split-coil magnet was used to allow the variation of the angle
between the plane of the sample and the field with a resolution
of $0.05^\circ$. Samples were mounted on a silicon-nitride
cantilever with length, width, and thickness of $210\mu m$, $50\mu
m$, and $0.51\mu m$ respectively. The cantilevers were made at
the Stanford Nanofabrication Facility employing standard
micromachining techniques.  Figure 1 shows a schematic of the
sample mounted on the cantilever, as well as the configuration of
the cantilever, sample, and magnetic field direction.

\begin{figure}
\includegraphics[width=0.8 \columnwidth]{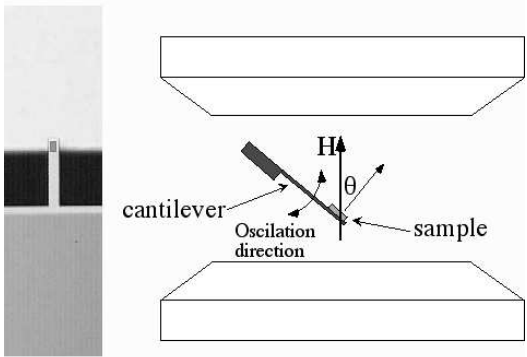}
\caption{Schematic of the experimental geometry. Theta ($\theta$)
is the angle between the field and the normal to the sample (and
to the  cantilever). At left is a top view of the silicon-nitride
cantilever with the sample outlined on its surface near the tip.}
\label{fig1}
\end{figure}

The fundamental frequency of the bare cantilever was  $f_0\sim 10
kHz$, yielding a spring constant $k\sim 0.02 N/m$. The quality
factor, $Q$, of the bare cantilever was measured to be $1.8
\times 10^4$ at room temperature and $8.0 \times 10^4$ at $4 K$.
Single crystals of BSCCO were grown by a directional
solidification method \cite{fournier} and their $T_c$ was
measured by SQUID magnetometry.  A slightly underdoped crystal
with $T_c$ of about $87.5 K$ was used. The crystal's dimensions
were $40\cdot 20\cdot 1\mu m^3$ and the sample was glued with a
thin layer of epoxy to the cantilever with the {\it c} axis
perpendicular to the cantilever surface. Though placing the
sample on the cantilever changed its natural frequency to $\sim
6500 Hz$ according to the additional mass (of sample and epoxy),
less than $20\%$ degradation in the $Q$ of the system was
observed due to the gluing process.

The cantilever displacement was determined interferometrically
using a fiber interferometer.  In such a device \cite{rugar},
light from a diode laser travels down a fiber situated just above
the the surface of the cantilever.  Light that bounces off the
cleaved end of the fiber interferes with light that exits the
fiber, reflects off the cantilever, and then re-enters the fiber.
The interferometric signal allows determination of the distance
between the fiber end and the cantilever with sub-angstrom
resolution for small measurement bandwidths.

The experimental procedure consisted of measurement of the
resonant frequency of the cantilever-sample system as a function
of temperature, magnetic field strength, and magnetic field
angle.  A self-oscillation drive system was designed to maintain
the cantilever vibration at a fixed amplitude.  In this system,
the thermal noise vibration signal from the cantilever is
amplified and shifted in phase, and then sent to a piezo actuator
that drives the cantilever. Because of the high quality factor of
these micro-machined cantilevers, the thermal noise induces
motion predominantly at the resonant frequency, and therefore,
the self-oscillation drive constantly maintains the system at
this resonant frequency.  The frequency is measured concurrently
with the amplitude of the drive signal necessary to sustain a
constant oscillation amplitude.  The amplitude of the drive
signal is inversely proportional to the $Q$ of the system, and is
hence a measure of the relative dissipation.

The gross behavior of the system of BSCCO crystal and cantilever
in perpendicular (field aligned with {\it c} axis and cantilever
normal) and parallel (field aligned with {\it ab} plane and long
axis of cantilever) field is depicted in figure 2. Here we plot
the resonant frequency of the system and its measured dissipation
as a function of magnetic field for both orientations. In
perpendicular field the dissipation is flat and translates to a
$Q$ of the combined system of $\sim 4.0 \times 10^4$ at $T\sim
T_c$. The frequency of the system changes by less than 0.01$\%$
of the fundamental frequency at $H=0$ in this field range. The
near-parallel behavior is very different, showing an increase in
the system's frequency ({\it i.e.} restoring force) with
increasing field; the dissipation peaks at $\sim 450 Oe$ in this
particular example. The meaning of the specific location of the
peak will be discussed later.

\begin{figure}
\includegraphics[width=0.85 \columnwidth]{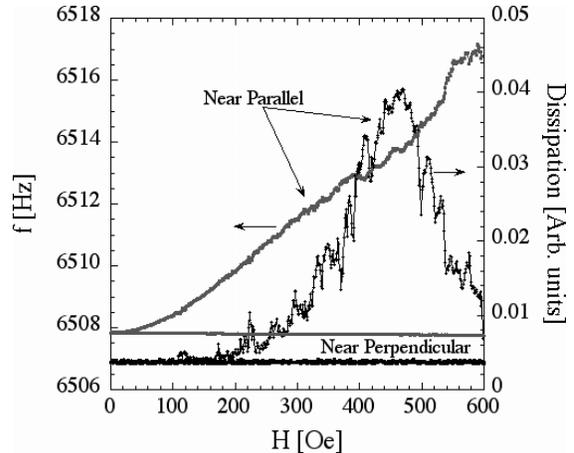}
\caption{Resonant frequency and dissipation $vs.$ magnetic field
($T=80 K$) for field orientations parallel and perpendicular to
the cantilever's long direction and simultaneously the sample's
{\it ab} plane.  } \label{fig2}
\end{figure}

\begin{figure}
\includegraphics[width=0.85 \columnwidth]{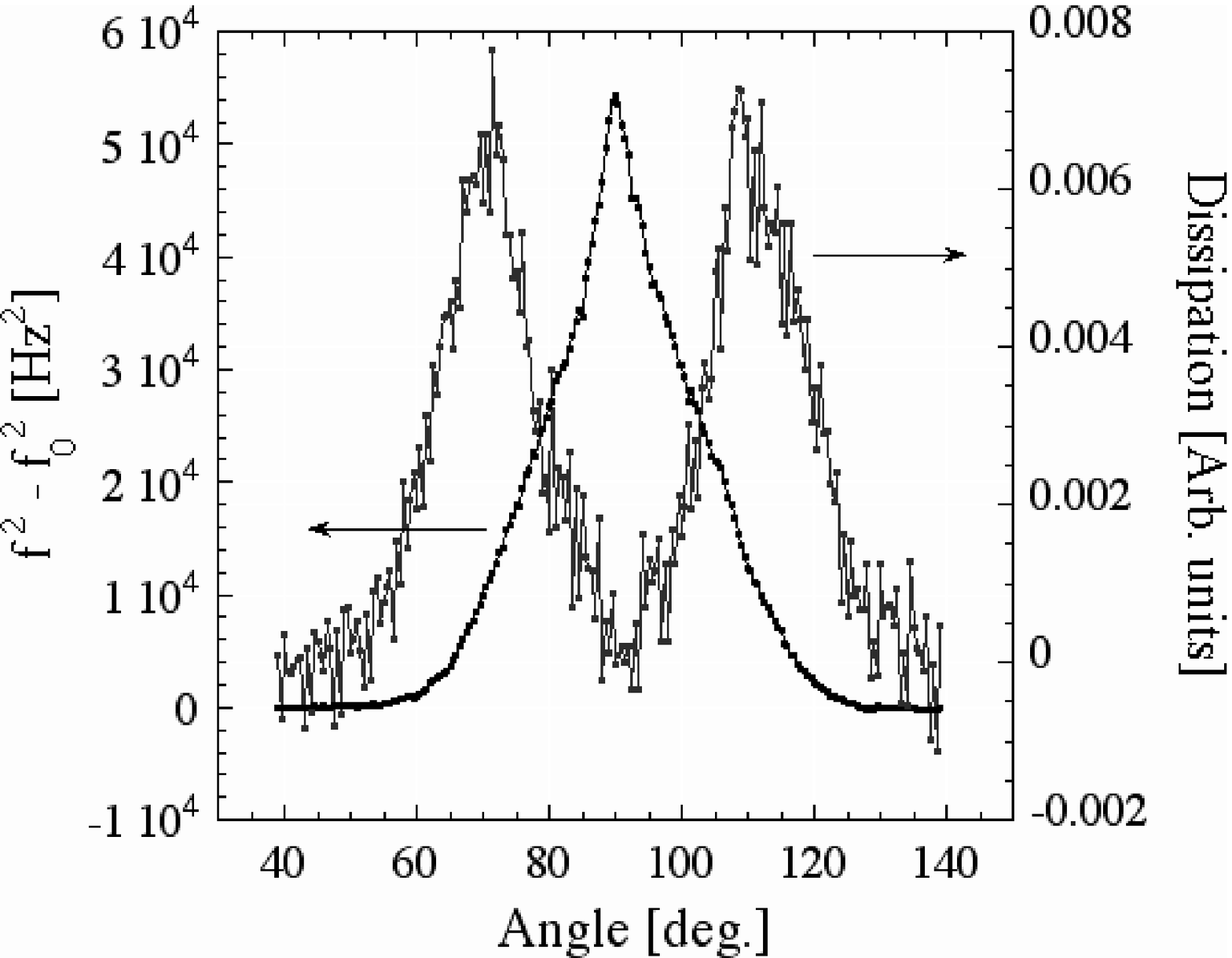}
\caption{Typical data from an angular scan ($T=75 K$, $H=160 Oe$).
Plotted are resonant frequency and dissipation $vs.$ the angle
between the crystal's {\it c} axis (as well as cantilever surface
normal) and the applied field. The resonant frequency peaks where
the field is aligned with the sample's Cu-O planes, while the
dissipation takes it's greatest value where the normal component
of the field induces a structural transition in the vortex
lattice, somewhat off parallel.  The curves are essentially
symmetric around the parallel configuration.} \label{fig3}
\end{figure}

Figure 3 depicts a typical scan of frequency shift and relative
dissipation of the cantilever and sample as a function of angle
at $T=75 K$ in a magnetic field of $160 Oe$. Note that the figure
is symmetric around $90^\circ$ where the field is parallel to the
sample's {\it ab} plane. The frequency shift measures the
restoring force of the magneto-mechanical system, while the
relative dissipation is indicative of the system's behavior as it
is altered from being more dissipative (small angles) to more
inductive (large angles). Such behavior is expected as
dissipation can occur only in the Cu-O planes (through motion of
pancake vortices). For small angles the field is almost
perpendicular to the Cu-O planes and thus its normal component
exceeds the melting field \cite{zeldov}. For angles closer to
$90^\circ$ the normal component of the field advances through the
melting point and the dissipative motion of free vortices changes
to a collective motion of the vortex lattice that provides an
extra restoring force to the magneto-mechanical system.

\begin{figure}
\includegraphics[width=0.8 \columnwidth]{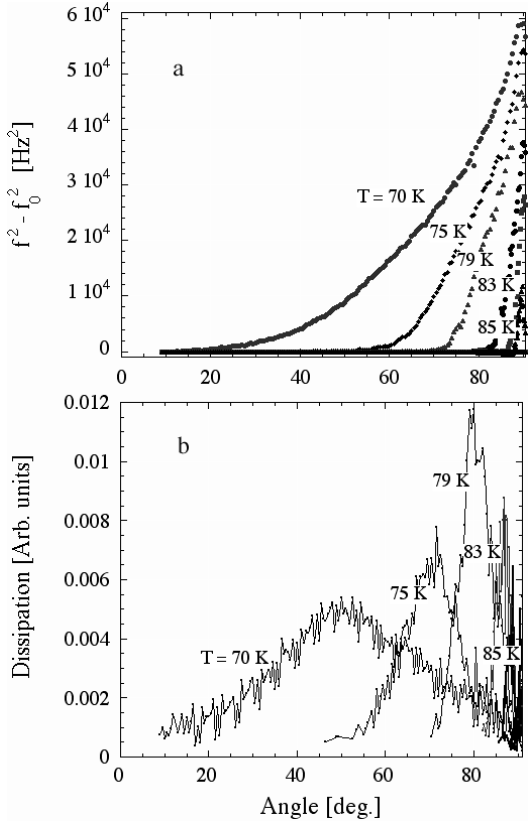}
\caption{Resonant frequency and dissipation as a function of
angle and temperature ($H=160 Oe$).  a) As $T$ approaches $T_c$,
the resonant frequency peak narrows, indicating a smaller
restoring force for angles away from parallel.  This is a
reflection of the fact that the melting field is higher at lower
temperature.   b) The position of the dissipation peak moves
toward higher angles as $T$ approaches $T_c$.  } \label{fig4}
\end{figure}

This general scenario is better described in figure 4. Figure 4a
shows the angular dependence of the frequency shift and figure 4b
the angular dependence of the relative dissipation for
temperatures in the range $70 K$ to $87 K$, just $0.5 K$ below
$T_c$. The nominal magnetic field is $160 Oe$. The increase in
frequency shift with increasing angle could simply indicate that
for a given temperature, above a certain angle, there is an added
restoring force due to the formation of a vortex lattice. However,
we also note from figure 2 that the equivalent measurement in
perpendicular field does not produce a similar increase in
resonant frequency. We are therefore led to conclude that there is
an additional magneto-mechanical coupling in the system that
increases with the change of angle towards parallel
configuration. The most likely explanation is the coupling of
Josephson vortices to the magnetic field, a subject that we will
discuss further below.

Concentrating first on the position of the peak as a function of
temperature (figure 4b) as an indication for dissociation of the
pancake vortex lattice, we show in figure 5 the respective
perpendicular component of the magnetic field $vs.$ temperature
for two different nominal magnetic fields.

\begin{figure}
\includegraphics[width=0.8 \columnwidth]{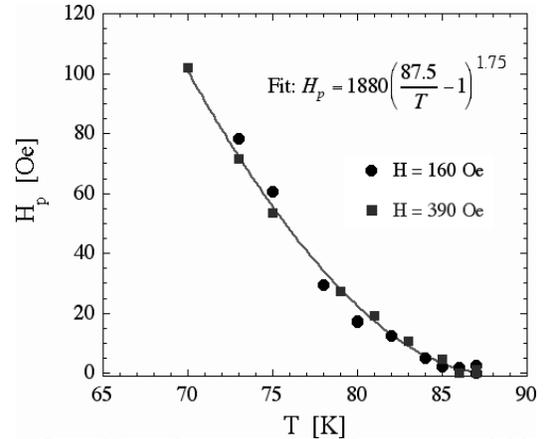}
\caption{Normal component of the magnetic field at the angular
location of the dissipation peaks as a function of temperature.
The line is a fit to a power law ($H_p \propto
(T_c(0)-T_c(H))^b$) with the best fit for an exponent of $b
=1.75$. }
\label{fig5}
\end{figure}

\noindent Fitting our data near the transition we find that $H_p
\propto (T_c(0) - T_c(H))^{b}$ with $b =1.75 \pm 0.1$. Comparing
our results to those of Zeldov {\it et al.} \cite{zeldov} who find
an exponent closer to $1.5$, we notice that our data exhibit more
curvature as a function of temperature. Note that quadratic
dependence is what one would expect based on a simple Lindeman
criterion calculation as was first calculated by Houghton {\it et
al.} \cite{houghton}. Our result is therefore in the range of
melting transition exponents, $b=1.3$ to $2.0$, cited in previous
work \cite{zeldov,schilling,revaz}.

One of the more interesting effects resulting from the Josephson
lattice is a shift in the melting point. Assuming negligible
entropic correction to the free energy of the Josephson lattice
and that the dominant coupling between layers is magnetic,
Koshelev \cite{koshelev1} calculated the slope of the linear
dependence of the melting fields in the perpendicular and
parallel directions.  Figure 6 shows typical data at $T=80 K$
where the perpendicular component ({\it i.e.} along the {\it c}
axis) of the field at the dissipation peak is plotted against the
parallel component ({\it i.e.} along the {\it ab} plane).

The linear fit shown in figure 6 indicates a slope of
$-0.07\pm0.01$. This value is in agreement with measurements on
BSCCO crystals performed at the same temperature ($T=80 K$) by Ooi
{\it et al.} \cite{ooi} and is very close to the value estimated
by Koshelev \cite{koshelev1}. Note that for moderately anisotropic superconductors
the perpendicular field depends quadratically on the in-plane
field. Thus, the linear dependence observed here is a direct
indication of crossing lattices, which should exist in highly
anisotropic superconductors like BSCCO.

\begin{figure}
\includegraphics[width=0.75 \columnwidth]{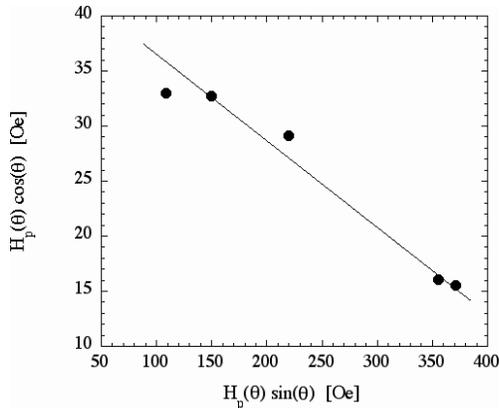}
\caption{Perpendicular component of field at dissipation peak
location $vs.$ parallel component of same.  The line is a linear
fit, after the prediction of Koshelev, with a slope of
$-0.07 \pm 0.01$ (See text.)}
\label{fig6}
\end{figure}

We finally discuss fine scans of the frequency shift near parallel
field. With the field magnitude fixed, the angle is varied around
parallel field. The data is shown in figure 7 below. Definite
oscillations are observed near parallel field and are
reproducible with a period of $\sim 0.4^\circ$ as derived from a
power spectrum of the data. The reproducibility is shown in
figure 8 where it can be seen that the peak positions do not
change between a sweep up in angle (magnetic field) and a
subsequent sweep down. In contrast, a similar experiment near
perpendicular magnetic field shows a flat response (see figure
7). Knowing the area of the sample in the parallel direction and
the applied magnetic field we find that the flux associated with
one period is $\sim(160Oe)\times (0.4\pi/180)\times (20\mu
m^2)\approx 2.2\times 10^{-7} Oe-cm^2 \approx \Phi_0$, where
$\Phi_0$ is one flux quantum. This result clearly indicates that
we are monitoring single Josephson vortices entering (or exiting)
the sample as the angle is changed.

\begin{figure}
\includegraphics[width=0.8 \columnwidth]{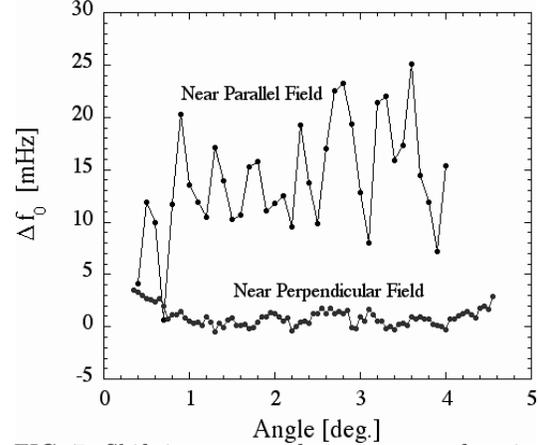}
\caption{Shift in resonant frequency as a function of angle
(values relative) around parallel configuration as compared to
the frequency shift in a similar range around perpendicular.  The
scatter in the latter data is on the level of the noise of the
measurement (a few $mHz$).  These data suggest the observance of
Josephson vortex motion.} 
\label{fig7}
\end{figure}

\begin{figure}
\includegraphics[width=0.8 \columnwidth]{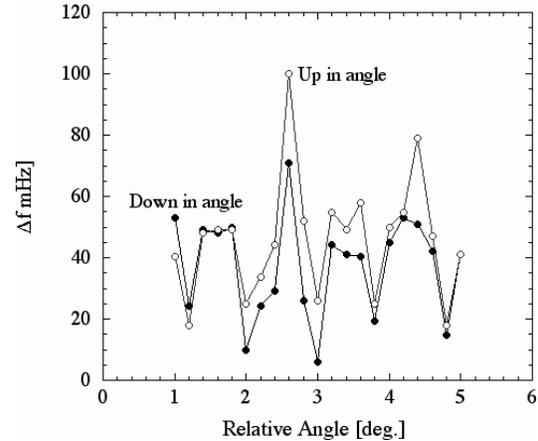}
\caption{Shift in resonant frequency $vs.$ angle around the
parallel configuration, displaying data for a sweep up in angle
as well as a sweep down.  The jumps are somewhat reproducible
with some hysteresis due to sample-specific vortex behavior.
Analysis of the power spectrum of this data implies a periodicity
corresponding to the entrance or exit of one Josephson vortex into
or out of the sample .} \label{fig8}
\end{figure}

In conclusion, we have presented a detailed study of the
low-field magnetic behavior of a nanogram-sized BSCCO single
crystal while varying the direction of the sample in the applied
field. Our data indicate a vortex structure consisting of two
vortex lattices. A pancake vortex lattice that melts according to
the perpendicular component of the magnetic field exists in the
{\it ab} planes. A Josephson vortex lattice penetrates the
material along the {\it ab} planes and melts at the same
temperature as the pancake vortex lattice. At angles very close
to parallel there exists only one row of Josephson vortices in
the {\it ab} planes, and upon changing the angle of the magnetic
field, single Josephson vortices are observed to enter or exit
the sample depending on the direction of the sweep of the field.

Work supported by DoE Grant DE-FG03-94ER45528. JC thanks DoD for
fellowship support. Samples prepared at Stanford's Center for
Materials Research. This work made use of the National
Nanofabrication Users Network facilities supported by the
National Science Foundation under Award ECS-9731294.

\end{document}